\documentclass[12pt,a4paper]{article}

\makeatletter 
\def\set@curr@file#1{%
  \begingroup
    \escapechar\m@ne
    \xdef\@curr@file{\expandafter\string\csname #1\endcsname}%
  \endgroup
}
\def\quote@name#1{"\quote@@name#1\@gobble""}
\def\quote@@name#1"{#1\quote@@name}
\def\unquote@name#1{\quote@@name#1\@gobble"}
\makeatother

\usepackage{graphicx}
\usepackage{xcolor}
\usepackage[normalem]{ulem}
\usepackage{url}
\usepackage{amssymb}
\usepackage{ulem}
\usepackage{authblk}
\usepackage[left=2.25cm, right=2.25cm, top=2cm, bottom=2.5cm]{geometry}

\date{}

\title{\textbf{Stress-strain in electron-beam activated polymeric micro-actuators}}

\author{
Davide Giambastiani$^{1,2}$, 
Fabio Dispinzeri$^{1,2}$, 
Francesco Colangelo$^{2}$\footnote{\textit{Permanent Address:} Institute for Quantum Electronics, ETH Z\"urich, CH-8093 Z\"urich, Switzerland}, Stiven Forti$^{3}$, Camilla Coletti$^{3,4}$,
Alessandro Tredicucci$^{1,2}$,
Alessandro Pitanti$^{2}$ and 
Stefano Roddaro$^{1,2}$
}

\affil{$^1$ Dipartimento di Fisica "E. Fermi", Università di Pisa, Largo B. Pontecorvo 3, I-56127 Pisa, Italy}
\affil{$^2$ NEST, CNR - Istituto Nanoscienze and Scuola Normale Superiore, piazza San Silvestro 12, I-56127 Pisa, Italy}
\affil{$^3$ Center for Nanotechnology Innovation @NEST, Istituto Italiano di Tecnologia, Piazza San Silvestro 12, I-56127 Pisa, Italy}
\affil{$^4$ Graphene Labs, Istituto Italiano di Tecnologia, Via Morego 30, I-16163 Genova, Italy}

\begin{document}

\maketitle

\begin{abstract}
Actuation of thin polymeric films via electron irradiation is a promising avenue to realize devices based on strain engineered two dimensional (2D) materials. Complex strain profiles demand a deep understanding of the mechanics of the polymeric layer under electron irradiation; in this article we report a detailed investigation on electron-induced stress on poly-methyl-methacrylate (PMMA) thin film material. After an assessment of stress values using a method based on dielectric cantilevers, we directly investigate the lateral shrinkage of PMMA patterns on epitaxial graphene, which reveals a universal behavior, independent of the electron acceleration energy. By knowing the stress-strain curve, we finally estimate an effective Young's modulus of PMMA on top of graphene which is a relevant parameter for PMMA based electron-beam lithography and strain engineering applications. 
\end{abstract}

\section{Introduction}

The mechanical deformation of polymers under electrostatic forces, electrostriction, ion insertion, and molecular conformation changes is a well known effect and can be exploited for creating polymeric artificial muscles~\cite{Mirfakhrai2007}. Recently, this concept has been shifted to nanoscale actuators and few experiments have demonstrated that 2D materials can be strained by simply releasing part of a stressed PMMA layer~\cite{Polyzos2015} or by acting on specific regions of the polymer using a focused electron beam (e-beam). In particular, e-beam induced polymerization of spin-on-glass material~\cite{Shioya2014} and shrinkage of PMMA~\cite{Colangelo2018, Colangelo2019} under large dose irradiation have been successfully used to control the local strain of graphene and other 2D materials.
This latter technique offers a key advantage: the deformation of the target layer can be controlled with a great freedom by defining a custom e-beam pattern, and thus by controlling the local stress induced in the PMMA actuator. Moreover, the spontaneous or heat-induced mechanical relaxation of the polymer over time allows to apply different strain profiles to the same device, in a sequence of actuation-relaxation cycles~\cite{Colangelo2018}. This is at odds with other techniques based on e-beam irradiation of metallic layers such as Ni~\cite{Shioya2014}, which instead give permanent strain profiles.

In a more standard context, the mechanical deformation of resists during lithographic exposure has been widely studied in the literature, with the aim of improving the resolution and precision of the features of the polymeric masks. A number of possible deformation mechanisms have been identified and investigated, including the thermal annealing of the resist due to electron irradiation~\cite{Habermas2002, Ayal2009}; the evaporation of trapped solvents~\cite{Kudo2001}; chemical reactions leading to chain scission and recombination during irradiation~\cite{Sarubbi2001, Azuma2004, Bunday2009, Yasuda2015}. Given the qualitatively different target of these studies, the polymeric layer typically displays a strong adhesion to the substrate, which is not ideal for the calibration of the large lateral deformations in our polymeric actuators. In addition, the investigation of resist thinning~\cite{Habermas2002, Sarubbi2001, Yasuda2015} -- although surely useful for lithography -- does not allow a reliable quantification of isotropic stress, which is a critical parameter to accurately design and predict the strain configuration induced by a planar actuator on a 2D material.

In this article, we quantitatively investigate the mechanical properties of PMMA resist under electron exposure. At first, we calibrate the PMMA stress from the mechanical response of PMMA/Si$_3$N$_4$ cantilevers after irradiation. We also independently perform a quantitative estimation of the in-plane deformation of a PMMA layer using a low-friction substrate consisting of epitaxial bilayer graphene (BLG) on SiC. Based on these experiments, we fully characterize the mechanical response of our actuators and we derive a PMMA effective Young's modulus of ($1.7\pm 0.1$) GPa, which is comparable with the values reported in the literature~\cite{Briscoe1998, Stafford2004, Yamazaki2007}. Our calibration is relevant not only for applications of polymer-based strain engineering, but also for any device whose behavior is affected by stress transfer at the interface between a polymeric layer and a 2D material~\cite{Gong2010, Jiang2014, Annagnostopoulos2015}. Applications go from the use of cross-linked PMMA as an anchorage for mechanical resonators~\cite{Weber2014} or for composite graphene/PMMA suspended membranes~\cite{Xu2020} up to stretchable/wearable devices~\cite{Akinwande2014, Jang2016, Jang2019}, where the shear forces and sliding between a 2D material and its soft substrate are particularly relevant.

\section{Results and discussion}

\subsection{Stress-dose calibration with cantilevers}\label{subsec1}
\label{sec_Cantilever}
\begin{figure*}[h!]
    \centering
    \includegraphics[scale=1.2]{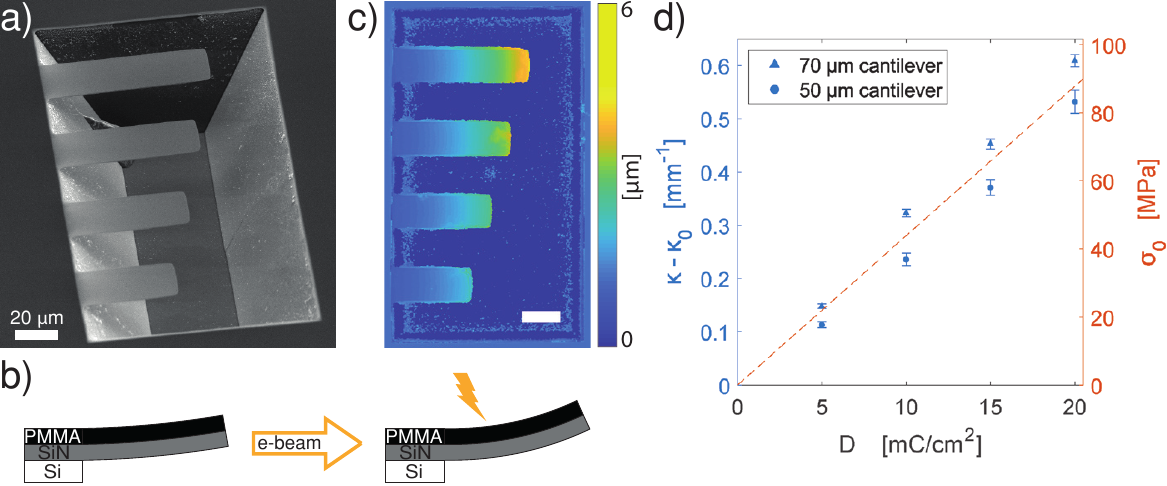}
    \caption{(a) SEM image of the cantilevers. (b) Sketch of the actuation mechanism: the cantilever has a different curvature before (left) and after (right) e-beam irradiation. (c) A typical height map taken at the optical profilometer (scale bar is $20\,\rm{\mu m}$). (d) Differential curvature $\kappa-\kappa_0$ for the $70\,\rm{\mu}$m (triangles) and $50\,\rm{\mu}$m (dots) cantilever as a function of the exposure doses (blue); curvature values are converted in stress values based on finite element simulations (orange dashed line, referring to the orange scale) after averaging the data for the two cantilever lengths.}
    \label{fig1}
\end{figure*}

The measurement of the deformation of a cantilever due to a thin overlayer is a well established method to measure the internal stresses stored in the coating material during the deposition process~\cite{Johansson1989, Ljungcrantz1993, Fang1995} or during thermal expansion~\cite{Chu1993, Fang1999}. By taking advantage of this technique, we measure the stress of a thin PMMA film as a function of the e-beam exposure parameters. The cantilevers are fabricated in an array of fourfold sets, with a fixed width of $20\,{\rm\mu m}$ and lengths in a range from $40$ to $70\,\rm{\mu m}$. A typical cantilever set is shown in the SEM picture in figure~\ref{fig1}(a). 
Spin-coating the polymeric layer on top of flexible, suspended cantilevers is not an ideal method for obtaining a thin and uniform PMMA film; we resorted to a transfer technique similarly to what routinely done with 2D materials~\cite{Van2016}. The fabrication procedure of the Si$_3$N$_4$-cantilevers is described in detail in the Methods section.
Once the devices have been fabricated, the calibration procedure follows the sketch of figure~\ref{fig1}(b): when the PMMA layer is exposed to e-beam radiation, the resulting rearrangement of the molecular structure and further effects tend to shift its mechanical equilibrium to a more compact configuration. This is typically modeled by introducing a ``pre-stress'' $\sigma_0>0$ in the constitutive relation $\sigma-\sigma_0 = C\!:\!\varepsilon$, where $C$ is the stiffness tensor. In the case of the PMMA-Si$_3$N$_4$ bilayer, mechanical equilibration leads to an upward bending of the cantilever. The measurement of the beam deformation by optical profilometry (see figure~\ref{fig1}(c)) allows to estimate $\sigma_0$, which is the key parameter for the design of the polymeric actuators.

Our analysis is based on the inverse bending radius of the cantilever, i.e. the curvature $\kappa$, which directly relates to $\sigma_0$. The value of $\kappa$ is extracted by fitting the height profiles using a second-order polynomial~\cite{Fang1995}. However, since the cantilever is already slightly bent upward due to the non-homogeneous tensile stress in Si$_3$N$_4$ along the growth direction~\cite{Shi2006}, we also measure the initial curvature $\kappa_0$ immediately before the PMMA actuation.  Subsequently, an additional tensile stress is applied by irradiating the PMMA with the e-beam at $5\,{\rm keV}$. After each exposure step, we measured the $50\,\rm{\mu}$m- and $70\,\rm{\mu}$m-cantilever curvature $\kappa$ and the resulting incremental curvature $\kappa-\kappa_0$, as reported in figure~\ref{fig1}(d).

In order to convert the measured curvature values into stress, we used a finite-element solver (COMSOL Multiphysics) and a cantilever model composed by two different layers with perfect adhesion (see Methods section). The bottom one is a $300\,{\rm nm}$-thick Si$_3$N$_4$ layer with Young's modulus $E_{SiN} = 250\,{\rm GPa}$ and Poisson ratio $\nu=0.23$. In the case of the top PMMA layer, we carefully took into account the thinning due to the e-beam irradiation, which is known to occur at low doses~\cite{Habermas2002, Sarubbi2001}. To this end, we measured the pristine and irradiated PMMA by Atomic Force Microscopy: the thickness of pristine PMMA was found to be $(97.7\pm1.2)\,{\rm nm}$ while we measured a thickness of $49-56\,{\rm nm}$ for the irradiated one, depending on the exact dose in the explored $5-20\,{\rm mC/cm^2}$ range. As argued in the following, the cantilever curvature is not particularly sensitive to the exact Young's modulus of the PMMA layer; differently, it is strongly sensitive on the value of the radiation-induced stress $\sigma_0$ of the polymeric layer. In particular, we verified that the relationship between $\sigma_0$ and $\kappa$ is linear and has a very weak dependence on the PMMA Young's modulus in the range $1-10\,{\rm GPa}$: $10\,{\rm GPa}$ in fact led to a slope of $\kappa(\sigma_0)$ that is only $2\%$ smaller than the one obtained for $1\,{\rm GPa}$ (for further discussion see the Methods section). This result agrees with previous numerical estimates~\cite{Mezin2006} where the residual stress of a thin film deposited on a plate was calculated from the curvature of the substrate, finding a weak dependence on the Young's modulus of the coating (see section \ref{subsec:fem}). As a result, we decided to use a standard $50\,{\rm nm}$ thickness and a PMMA Young's modulus of $1\,{\rm GPa}$ in all our simulations. The resulting conversion between curvature and stress values is visible in the right axis of figure~\ref{fig1}(d), where we averaged the data measured for the 50 and 70 $\rm \mu$m cantilevers since the simulated curvature as a function of the stress is essentially independent from the cantilever length. For the studied dose range, we observe that the estimated stress is proportional to the exposure dose $D$ and the stress-dose data can be satisfactorily fitted by a linear function $\sigma_0 = \alpha D$ with 

\begin{equation}
\alpha = (440 \pm 36)\,{\rm N/mC}.
\label{ParamCalib}
\end{equation}

We note that the resulting experimental uncertainty ($\approx 6\%$) is substantially larger than the simulation error deriving from our lack of knowledge of the exact value of the Young's modulus of the PMMA overlayer.

\subsection{Strain-dose calibration on epitaxial graphene}\label{subsec2}
\label{sec_Cal}
\begin{figure}[h!]
    \centering
    \includegraphics[scale=0.8]{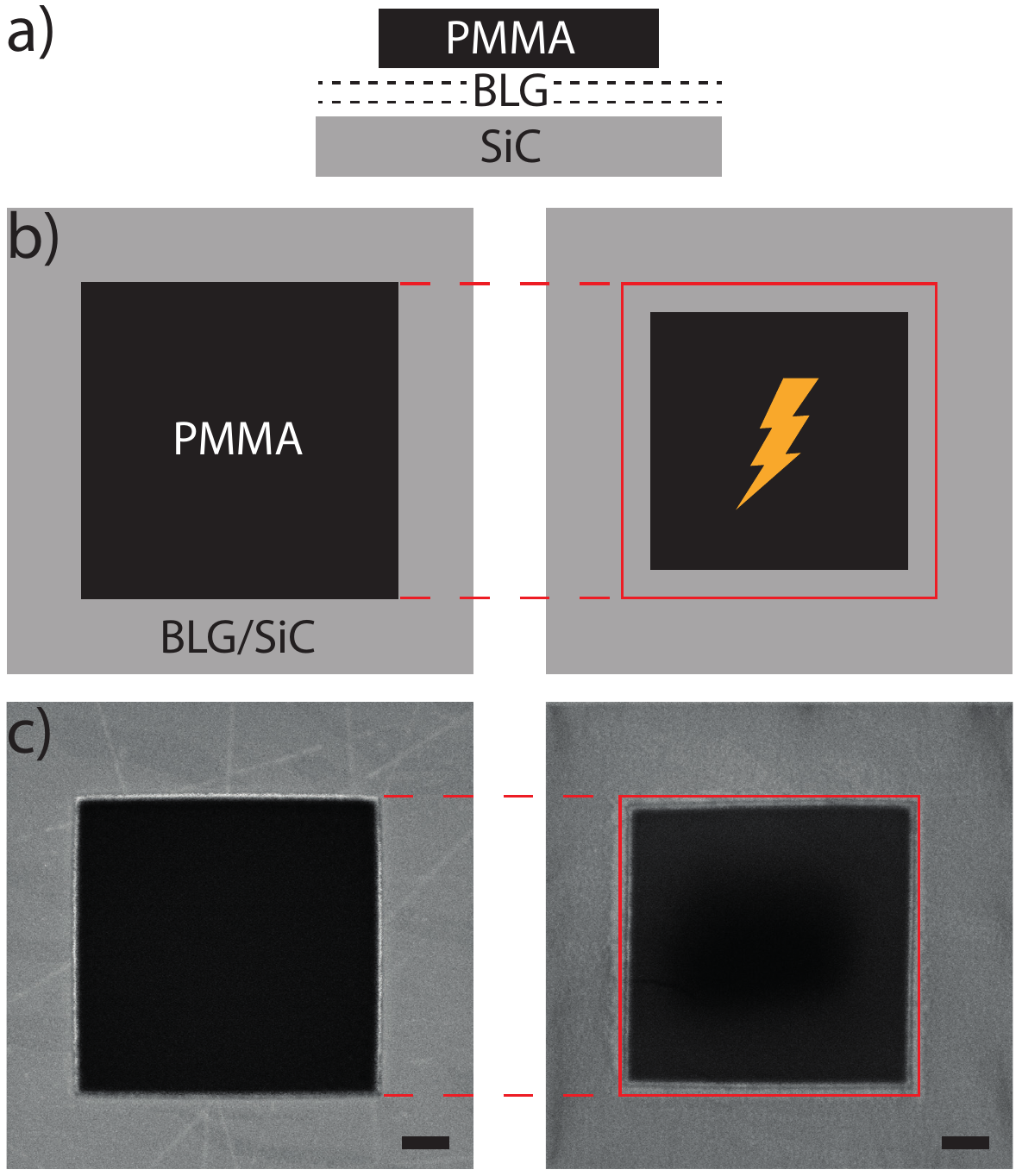}
    \caption{(a) Schematic cross section of the device under investigation with the bilayer epitaxial graphene (BLG) grown on SiC and a patterned square of PMMA. (b) Schematic of the actuation process: on the left, the square before and, on the right, the square after the e-beam irradiation. (c) First (left) and last (right) SEM image of one of the patterned squares in a sequence of irradiation steps  at $2\,{\rm keV}$. The dark square is PMMA while the bright region is the epitaxial graphene. The black bars are $600\,{\rm nm}$.}
    \label{fig2}
\end{figure}
Since the value of the Young's modulus is an important parameter to predict the behavior of the actuators, we performed a complementary experiment where we estimated the PMMA lateral shrinkage as a function of e-beam energy and dose. To this end, a set of PMMA patterns were fabricated on top of epitaxial bilayer graphene (BLG) grown on SiC (0001)~\cite{Emtsev2009, Riedl2007, Emtsev2008, Goler2013, Forti2014}. The reason for this choice is twofold: (i) this experimental configuration is relevant for other strain-engineering experiments~\cite{Colangelo2019}; (ii) we expect a relatively small friction and thus an easy lateral shrinkage of PMMA, given the flatness of the BLG/SiC substrate~\cite{First2010} and based on the results for adhesion at polymer/graphene interfaces \cite{Gong2010, Jiang2014, Annagnostopoulos2015}. The PMMA layer was spin-coated at $2000\,{\rm rpm}$ for 1 minute, baked at 90$^{\circ}$C for 15 minutes and patterned by standard e-beam lithography. 
The cross section of the system is sketched in figure~\ref{fig2}(a).
The basic experimental procedure is sketched in figure~\ref{fig2}(b): in a sequence of exposure steps, we first irradiate the PMMA with the e-beam and then measure the resulting shrinkage. Since SEM imaging already leads to a well-defined exposure of the PMMA, the measurement simply consists of a continuous image scanning of the polymer. The advantage of this technique is that PMMA can be actuated and probed very quickly and while keeping a steady environment, minimizing natural mechanical relaxation. In figure \ref{fig2}(c) are depicted the first (left) and last (right) SEM images of one of the patterned squares over the BLG/SiC substrate in a sequence of exposure steps.

\begin{figure}
    \centering
    \includegraphics[scale=0.9]{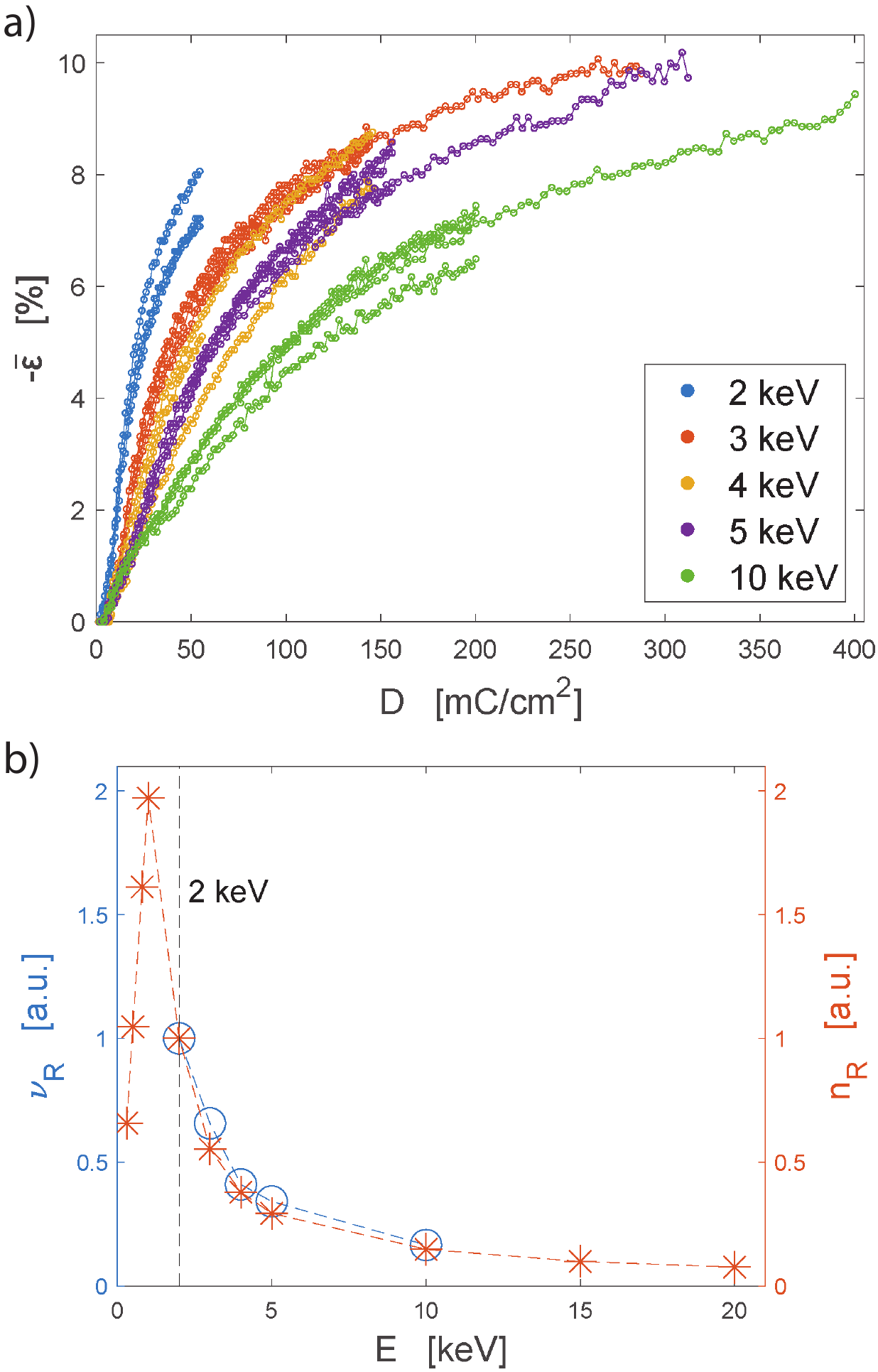}
    \caption{(a) Measured average planar strain $\overline{\varepsilon}$ of the PMMA squares as a function of the electron dose $D$ and beam energy $E$. Different curves of the same color represent different squares analysed, at the corresponding energy. (b) Dependence of the slope, $\rm{\nu_R}$, on the electron-energy (blue) and simulated number of electron-atom collisions, $\rm{n_R}$ (orange).}
    \label{fig3}
\end{figure}

In each measurement, we estimated the contraction of PMMA along the two sides of the square based on the contrast profile in the SEM image and calculated the average planar strain $\overline{\varepsilon}$. An evident PMMA shrinkage up to about $10\%$ is observed; the resulting trend is displayed in figure~\ref{fig3}(a) as a function of electron dose $D$ and at a set of different electron energies $E$ relevant for actuators, in a range from $E=2$ to $10\,{\rm keV}$. All the curves share a qualitative similar trend: (i) in the typical operational range of our actuators, the shrinkage initially displays a linear dependence on the dose with a slope $\nu=d\overline{\varepsilon}/dD$; (ii) beyond a given threshold, shrinkage tends to saturate. Both the slope of the linear region $\nu$ and the saturation threshold are found to strongly depend on the beam energy $E$ and for every given dose we observe a larger shrinkage at smaller energies.

A dependence of the shrinkage on the electron energy is indeed not unexpected, because the number of collisions between the electrons and the PMMA molecules depends strongly on the beam energy. In fact, PMMA is completely opaque to electrons at low energy, while it becomes almost transparent at high energies where most of the scattering occurs in the underlying substrate. In particular, collision events are expected to decay once the penetration depth exceeds the PMMA thickness.
\begin{figure}
    \centering
    \includegraphics[scale=0.9]{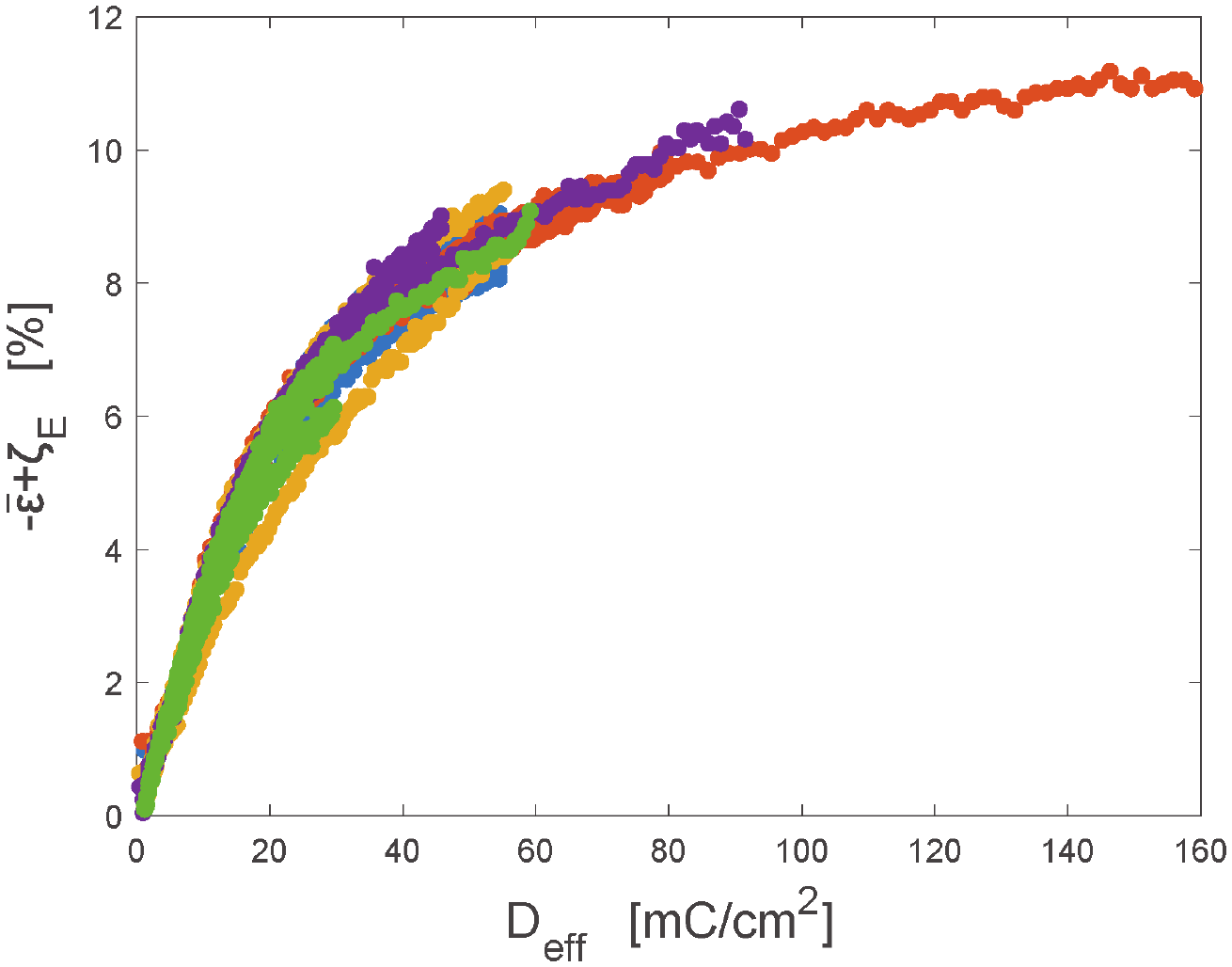}
    \caption{Universal trend of the shrinkages, translated by $\rm{\zeta_E}$ (see the Methods section), as a function of the effective dose. The color code is the same of that in figure~\ref{fig3}.}
    \label{fig4}
\end{figure}
In order to investigate how the electron energy affects the shrinkage of the PMMA, we performed a Monte-Carlo simulation of the collision of $10^4$ electrons with a $50\,{\rm nm}$ layer of PMMA; the estimate has been done using a freely available software~\cite{Drouin2007, Demers2011}. The number of collisions $n_{\rm coll}(E)$ has been calculated for several energies $E$, neglecting effects due to reflections of electrons from the substrate and secondary electron generation. The relative number of collisions $n_R = n_{\rm coll}(E)/n_{\rm coll}(2\,{\rm keV})$, defined using the $2\,{\rm keV}$ case as a reference, is reported in figure~\ref{fig3}(b) (orange line). Below $2\,{\rm keV}$, the number increases with energy and reaches a maximum around $1\,{\rm keV}$, while a marked decrease is observed at larger energies due to the increased penetration depth of the e-beam. A comparison between the relative number of electron-atom collisions and the observed PMMA shrinkage clearly shows that the simulated effect largely explains the energy dependence we observe in our data. This can be easily seen in figure~\ref{fig3}(b): the relative slope $\nu_R = \nu(E)/\nu(2\,{\rm keV})$ (blue line) matches very nicely with the results of the Monte Carlo simulation (orange line). Based on this observation we define an effective electron dose $D_{\rm e\!f\!f}$, again using $2\,{\rm keV}$ as a reference,

\begin{equation}
    D_{\rm e\!f\!f} = \frac{D(E)\cdot n_{\rm coll}(E)}{ n_{\rm coll}(2\,{\rm keV})}
\end{equation}
and obtain that all the shrinkage-dose curves roughly collapse on a unique universal behavior, independent on the electron energy employed in the experiment (see figure~\ref{fig4}). We note that the regime we investigate is different from what reported in other lower energy studies~\cite{Kudo2001, Sarubbi2001, Habermas2002}, where the penetration depth is smaller than the PMMA thickness and larger shrinkage is observed at larger energy. This regime is not ideal for actuators, since it is important to include a uniform shrinkage throughout the whole PMMA thickness, thus the penetration depth has to be large. 

One interesting aspect of our experiment consists in the relative friction between PMMA and the underlying epitaxial graphene, which here we assumed to be negligible. On the other hand, it is also known that, upon exposure, the resist can display some interlayer adhesion and be used to strain an underlying 2D material~\cite{Colangelo2018, Colangelo2019}. The exact mechanisms leading to adhesion are still under investigation, but our previous works showed that beam irradiation is a crucial factor and increases adhesion, compared to that observed for unexposed PMMA~\cite{Colangelo2018}. Adhesion is also likely to be promoted by a rough interface topography but, since BLG/SiC is atomically flat, such effects should be minimized in our experimental configuration.

\subsection{Combined analysis}

\begin{figure}
    \centering
    \includegraphics[scale=0.9]{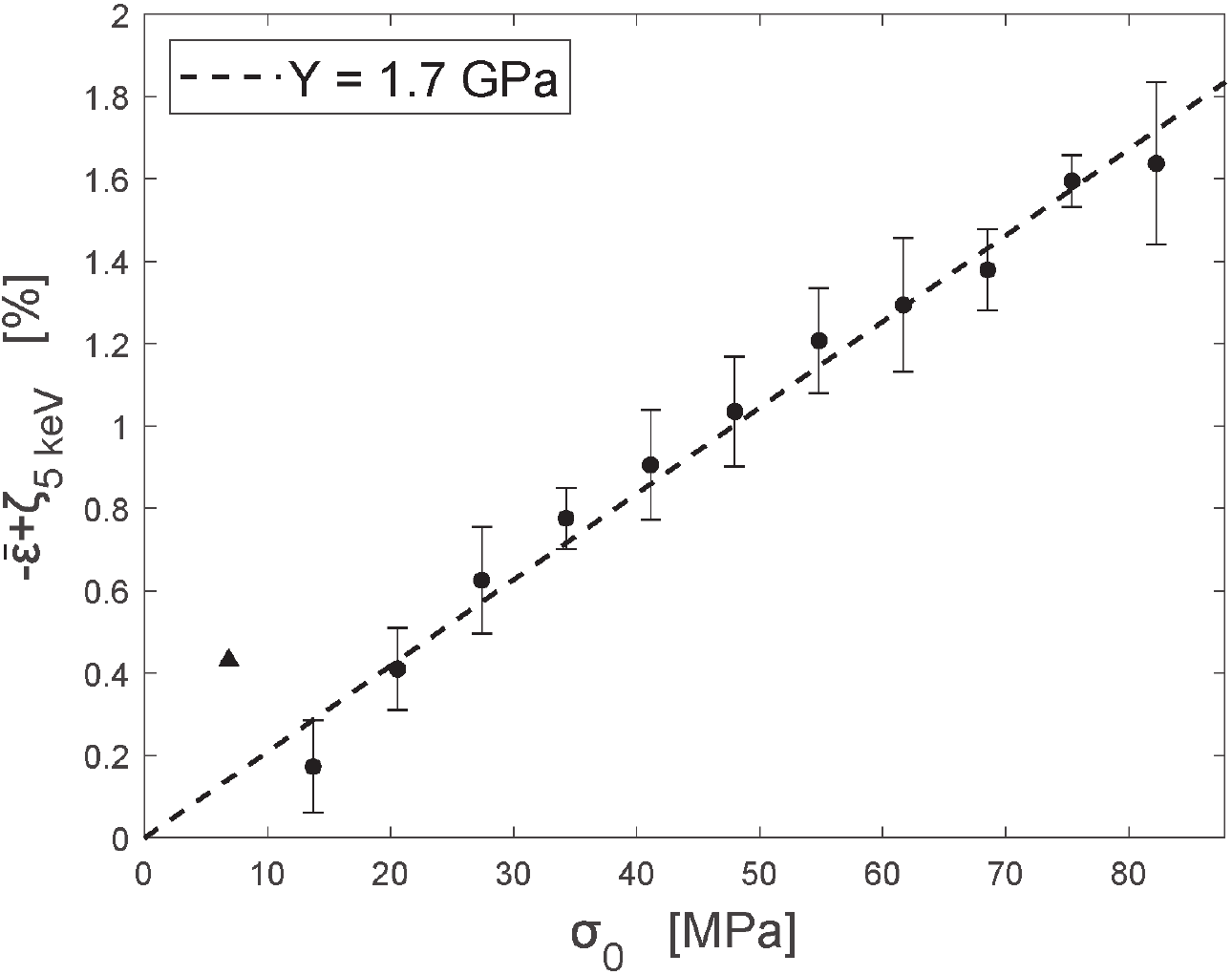}
    \caption{Average values of the shrinkage for 3 different PMMA squares, at each value of dose, as a function of the simulated stress calibrated with equation~(\ref{ParamCalib}). The error-bars are the standard deviation of the measured PMMA-squares. The triangle is defined to be the reference value ($\overline{\varepsilon}=0$). The dashed line is the numerical fit of data from which we estimate the Young's modulus Y.}
    \label{fig5}
\end{figure}

Results from the previous two sections can be combined to extract an estimate of the expected free shrinkage of PMMA as a function of the applied stress, in the linear actuation regime. This allows assessing the mechanical properties of our PMMA. In figure~\ref{fig5} we report data obtained by averaging the results of shrinkage experiments obtained on three different PMMA squares at $5\,{\rm keV}$. In the plot, the dose $D$ has been converted into a PMMA stress $\sigma_0$ based on the results of the cantilever calibration. Using the stress-strain relation for a homogeneous stress~\cite{Landau1986}, we fit the data and estimate an effective Young's modulus of ($1.7\pm0.1$) GPa for the e-beam actuated PMMA. Despite the measured PMMA Young's modulus could be affected by the chemical modifications and heating during e-beam exposure, are values are well within the range between $1$ and $3\,{\rm GPa}$ obtained  by Yamazaki \textit{et al.}~\cite{Yamazaki2007}.

\section{Conclusion}

In conclusion, we studied the mechanical response of PMMA under e-beam irradiation to extract useful parameters for the design of PMMA-based actuators. This was accomplished using two complementary methods: in a first set of experiments, PMMA was deposited on Si$_3$N$_4$-cantilevers and the stress vs dose response was evaluated based on the mechanical deflection of the cantilever and on finite element simulations; in a second set of experiments, the free contraction of PMMA patterned onto an epitaxial BLG was investigated. Our study highlights that the strong energy dependence we observe in the mechanical response above $2\,{\rm keV}$ can be largely explained in terms of an energy-dependent number of electron collision in the PMMA. As a result, we found an universal shrinkage-dose curve that is valid in the whole energy range relevant to the actuators. Finally, from the comparison of these two results we estimate the effective Young's modulus of PMMA patterned onto epitaxial BLG. Our work provides a benchmark for future applications of PMMA as an e-beam actuating layer for strain-engineering of 2D materials and for investigating the mechanical response of other polymeric thin films.

\section*{Acknowledgments}

This work has been financially supported by the Italian Ministry of Foreign Affairs (QUANTRA project) and the Italian Ministry of University and Reasearch (PRIN project QUANTUM2D and PRIN project MONSTRE-2D). This project has received funding from the European Union’s Horizon 2020 research and innovation programme Graphene Flagship under grant agreement No 881603.

\section{Methods}
\label{methods}

\subsection{\rm Si\textsubscript{3}N\textsubscript{4}-cantilever fabrication and PMMA transfer}

The substrate consists of a $250\,\rm{\mu m}$ commercial silicon wafer sandwiched between two $300\,{\rm nm}$ thick stoichiometric silicon nitride layers deposited with PECVD at high temperature ($\approx 850\,{\rm ^{\circ}C}$) in order to have a strong residual tensile stress at room temperature. For the definition of the sets of cantilevers, S1818 was spin-coated on the sample at $4000\,{\rm rpm}$ for 1 minute. After 1 minute of bake at $90\,{\rm ^{\circ}C}$ and pre-develop in MF-319 and H$_2$O, we patterned the sets of cantilevers through laser writer lithography exposing with a wavelength of $385\,{\rm nm}$ and energy dose of 150 mJ/cm$^2$. The sample was then baked 20 seconds at $120\,{\rm ^{\circ}C}$ and the pattern was developed in MF-319 and H$_2$O.
The S1818 was used as a mask for CF$_4$-based plasma etching. Then, wet etching of Si (KOH:H$_2$O 30$\%$ at $70\,{\rm ^{\circ}C}$) was performed. The resulting trench in the Si substrate is designed to be deep enough so that the device can survive wet processing and the resulting capillarity forces, which are relevant during the drying phase.

The preparation of PMMA (AR-P 679.02) for the transfer consisted of spin-coating in two successive steps, PVA and PMMA, both at $2000\,{\rm rpm}$ for 1 minute and baked at $90\,{\rm ^{\circ}C}$ for 1 minute, on silicon. Finally, the PMMA was removed from the substrate placing the sample in H$_2$O at 90$^{\circ}$C and was transferred onto the cantilevers and baked at $110\,{\rm ^{\circ}C}$ for 1 minute. The suspended PMMA was patterned by e-beam lithography and developed in a solution of Methyl Isobuthyl Ketone and Isopropyl alcohol (1:3) for 2 minutes and rinsed in Isopropyl alcohol in order to remove the portions of PMMA outside the cantilevers. In order to avoid charging effects, a conductive layer (ESPACER) is also used to coat the devices before the actuation experiment.

The profiles of the cantilevers were taken before and after electron irradiation at 5 keV, using a commercial optical profilometer (Taylor-Hobson, CCI HD).

\subsection{Analysis of the SEM profiles}
The size of the PMMA squares is measured extracting the profiles from each SEM-image ($1200\times1200$ pixels) which extends over an area of $6\times6$ $\mu$m. In order to reduce the noise, we performed an average of contiguous line-profiles corresponding to 0.4 $\mu$m and calculated the vertical and horizontal size of the PMMA square. From the analysis, we found that the vertical size of the PMMA is slightly larger than the horizontal one. The initial discrepancy of about 2.9$\%$ tends to decrase during e-beam irradiation and reaches about 2.5$\%$ at large electron doses. Regarding the analysis of the shrinkage during e-beam irradiation, we found that the first exposure apparently expands the PMMA ($\overline{\varepsilon}>0$). This effects is visible only for energies smaller than 10 keV and is probably due to charging of PMMA. In order to correct this systematic effect, we fitted the linear trend at low-doses (for each energy) with:
\begin{equation}
-\overline{\varepsilon} = \nu \cdot D - \zeta,
\end{equation}
where $\nu$ is defined in section~\ref{sec_Cal} and $\zeta$ is the offset.

\subsection{Finite element simulations}\label{subsec:fem}
The Stoney formula~\cite{Stoney1909, Janssen2009} gives the curvature $\tilde{\kappa}$ of a plate as a function of the planar stress of a thin deposited film $\sigma_f$:
\begin{equation}
\sigma_f =\frac{t_s^2 Y_s}{6t_f(1-\nu_s)}\tilde{\kappa},
\label{eq:stoney}
\end{equation}
where $t$, $Y$ and $\nu$ are thickness, Young's modulus and Poisson's ratio of the substrate (indicated with the subscript $s$) and the thin film (subscript $f$), respectively.
If the thin film thickness is comparable to the substrate one, the formula can be corrected with an extra term which depends linearly on the ratio between the two thicknesses $t_f/t_s$~\cite{Mezin2006}: \begin{equation}
\sigma_f =\frac{t_s^2 Y_s}{6t_f(1-\nu_s)} \tilde{\kappa}\times \nonumber \left[1+\left( 4\frac{Y_f(1-\nu_s)}{Y_s(1-\nu_f)}-1 \right)\frac{t_f}{t_s} \right],
\label{Stoney_Thick}
\end{equation}
Even if Stoney's formula was obtained for unclamped plates, as a first approximation it can be used to describe the curvature of clamped cantilevers. Therefore, we compared the analytical results of equation~(\ref{Stoney_Thick}) with finite element simulations of an initially flat PMMA/Si$_3$N$_4$ cantilever subjected to a planar stress of the PMMA layer. We found a good agreement between analytical results and simulations, with a maximum discrepancy of the ratio $\tilde{\kappa}/\sigma_f$ of about 9$\%$, considering the Young's modulus of PMMA $Y_f$ in a range from $1$ to $10\,{\rm GPa}$. Also, as previously discussed, equation~(\ref{Stoney_Thick}) shows a weak dependence from $Y_f$, with a few percent difference when the two extremal values in the considered range are used. The comparison between equation~(\ref{Stoney_Thick}) (which successfully reproduced numerical results~\cite{Mezin2006}) and our numerical simulations of flat cantilevers, validates our model, enabling its use for more complex geometries (pre-bended cantilevers), as the ones investigated in our experiment.

Finally, we note that a fundamental assumption of the model is that the two layers in the beam display a perfect adhesion. Indeed, we observed PMMA sliding only on top of BLG/SiC substrates, while no measurable sliding could be observed in similar experiments performed on SiC nor on ${\rm Si_3N_4}$.

\subsection{Epitaxial graphene on Silicon Carbide}
Epitaxial BLG was grown starting from nominally on-axis Si-face polished 6H-SiC(0001) purchased from SiCrystal Gmb. The synthesis of BLG was carried out by thermal decomposition, adapting the recipe of Emtsev and coworkers~\cite{Emtsev2009} in an Aixtron Black Magic reaction chamber, using an Ar backpressure of 780 mbar and a sublimation temperature of 1395 °C, measured with a single-wavelength infrared pyrometer focussed on the sample.

\bibliographystyle{BSTart}
\bibliography{References}

\end{document}